\documentclass[acus]{JAC2001}

\usepackage{graphicx}

\begin{document}

\title{MACROSCOPIC FLUID APPROACH TO THE COHERENT \\ 
BEAM-BEAM INTERACTION} 

\author{Stephan I. Tzenov and Ronald C. Davidson\\ 
Plasma Physics Laboratory, Princeton University, Princeton, 
New Jersey 08543, USA}

\maketitle

\*\vspace{0.0cm}

\begin{abstract}
Building on the Radon transform of the Vlasov-Poisson equations, a 
macroscopic fluid model for the coherent beam-beam interaction has 
been developed. It is shown that the Vlasov equation, expressed in 
action-angle variables, can be reduced to a closed set of 
hydrodynamic (fluid) equations for the beam density and current 
velocity. The linearized one-dimensional equations have been 
analysed, and explicit expressions for the coherent beam-beam 
tuneshifts are presented. 
\end{abstract}

\renewcommand{\theequation}{\thesection.\arabic{equation}}

\setcounter{equation}{0}

\*\vspace{0.0cm}

\section{INTRODUCTION}

In a colliding-beam storage ring device, the evolution of each beam 
is strongly affected by the electromagnetic force produced by the 
counter-propagating beam. A basic feature of this coherent 
interaction is linear mode coupling, also known as the coherent 
beam-beam resonance.

The problem of coherent beam-beam resonances in one dimension (the 
vertical direction) was first studied by Chao and Ruth \cite{chao} 
by solving the linearized Vlasov-Poisson equations. They considered 
the simplest case of a symmetric collider and obtained explicit 
expressions for the resonance stopbands. The purpose of the present 
paper is to extend their results to the case of an asymmetric 
circular collider. 

Based on the Radon transform \cite{deans,tzenov}, a macroscopic 
fluid model of the coherent beam-beam interaction is developed. The 
linearized macroscopic fluid equations are then solved, and a 
generalized stability criterion for a coherent beam-beam resonance 
of arbitrary order is derived. 

\renewcommand{\theequation}{\thesection.\arabic{equation}}

\setcounter{equation}{0}

\section{THE RADON TRANSFORM}

We begin with the one-dimensional Vlasov-Poisson equations describing 
the nonlinear evolution of the beams in the vertical $(y)$ direction 
\begin{equation}\label{eq:vlasov} 
{\frac {\partial f_k} {\partial \theta}} + 
\nu_k p {\frac {\partial f_k} {\partial y}} - 
{\frac {\partial {\cal H}_k} {\partial y}} 
{\frac {\partial f_k} {\partial p}} = 0, 
\end{equation} 
\begin{equation}\label{eq:hamilt} 
{\cal H}_k = {\frac {\nu_k} {2}} {\left( 
p^2 + y^2 \right)} + \lambda_k \delta_p (\theta) 
V_k(y; \theta), 
\end{equation} 
\begin{equation}\label{eq:poisson} 
{\frac {\partial^2 V_k} {\partial y^2}} = 4 \pi 
\int dp f_{3-k}(y, p; \theta), 
\end{equation} 
\begin{equation}\label{eq:lambda} 
\lambda_k = {\frac {R r_e N_{3-k} \beta_{ky}^{\ast}} 
{\gamma_{k0} L_{(3-k)x}}} {\frac {1 + \beta_{k0} 
\beta_{(3-k)0}} {\beta_{k0}^2}} \simeq 
{\frac {2 R r_e N_{3-k} \beta_{ky}^{\ast}} 
{\gamma_{k0} L_{(3-k)x}}}. 
\end{equation} 

\*\vspace{0.2cm}

\noindent 
Here, $(k = 1, 2)$ labels the beam, $f_k(y, p; \theta)$ is the 
distribution function, $\theta$ is the azimuthal angle, $\nu_k$ is 
the betatron tune in vertical direction, $R$ is the mean machine 
radius, $r_e$ is the classical electron radius, $N_{1,2}$ is the 
total number of particles in either beam, $V_k(y; \theta)$ is the 
normalized beam-beam potential, $\beta_{ky}^{\ast}$ is the vertical 
beta-function at the interaction point, and $ L_{kx}$ is the 
horizontal dimension of the beam ribbon \cite{chao}. The 
one-dimensional Poisson equation (\ref{eq:poisson}) can be readily 
solved to give 
\begin{equation}\label{eq:potent} 
V_k(y; \theta) = 2 \pi \int dy' dp' f_{3-k}{\left( 
y', p'; \theta \right)} {\left| y - y' \right|}. 
\end{equation} 
\noindent 
Transforming to action-angle variables $(J, \varphi)$, we rewrite 
Eqs. (\ref{eq:vlasov}) and (\ref{eq:hamilt}) in the form 
\begin{eqnarray}\label{eq:vlasaa} 
{\frac {\partial f_k} {\partial \theta}} + 
{\frac {\partial} {\partial \varphi}} 
{\left[ {\left( \nu_k + \lambda_k \delta_p(\theta) 
{\frac {\partial V_k} {\partial J}} \right)} f_k 
\right]} \nonumber \\ 
- {\frac {\partial} {\partial J}} 
{\left( \lambda_k \delta_p(\theta) 
{\frac {\partial V_k} {\partial \varphi}} f_k 
\right)} = 0, 
\end{eqnarray} 
\begin{equation}\label{eq:hamilaa} 
{\cal H}_k = \nu_k J + \lambda_k \delta_p(\theta) 
V_k(\varphi, J; \theta), 
\end{equation} 
\noindent 
where 
\begin{eqnarray}\label{eq:potenaa} 
V_k(\varphi, J; \theta) = 2 \pi \int d \varphi' 
dJ' f_{3-k}{\left( \varphi', J'; \theta 
\right)} \nonumber \\ 
\times {\left| \sqrt{2J} \cos \varphi - \sqrt{2J'} 
\cos \varphi' \right|}. 
\end{eqnarray} 
\noindent 
Next we perform the Radon transform defined as \cite{deans,tzenov} 
\begin{equation}\label{eq:radon} 
f_k(\varphi, J; \theta) = \int d \xi 
\varrho_k(\varphi, \xi; \theta) \delta {\left[ 
J - v_k(\varphi, \xi; \theta) \right]}, 
\end{equation} 
\noindent 
and obtain the hydrodynamic equations 
\begin{equation}\label{eq:hydrho} 
{\frac {\partial \varrho_k} {\partial \theta}} + 
{\frac {\partial} {\partial \varphi}} 
{\left[ {\left( \nu_k + \lambda_k \delta_p(\theta) 
{\frac {\partial V_k} {\partial v_k}} \right)} 
\varrho_k \right]} = 0, 
\end{equation} 
\begin{eqnarray}\label{eq:hydrov} 
{\frac {\partial {\left( \varrho_k v_k \right)}} 
{\partial \theta}} + 
{\frac {\partial} {\partial \varphi}} 
{\left[ {\left( \nu_k + \lambda_k \delta_p(\theta) 
{\frac {\partial V_k} {\partial v_k}} \right)} 
\varrho_k v_k  \right]} \nonumber \\ 
+ \lambda_k \delta_p(\theta) 
{\frac {\partial V_k} {\partial \varphi}} 
\varrho_k = 0, 
\end{eqnarray} 
\noindent 
where $\varrho_k$ is the Radon image of the distribution function 
$f_k$. The integration variable $\xi$ is regarded as a Lagrange 
variable, that keeps track of the detailed information about the 
action $J$. It us usually determined by the condition that the 
distribution function $f_k$ be equal to a specified distribution 
\cite{tzenov}, from which $J = v_k(\varphi, \xi; \theta)$. Taking 
into account Eq. (\ref{eq:hydrho}), the beam density can be further 
eliminated from Eq. (\ref{eq:hydrov}), which yields the result 
\begin{equation}\label{eq:hydvel} 
{\frac {\partial v_k} {\partial \theta}} + 
{\left( \nu_k + \lambda_k \delta_p(\theta) 
{\frac {\partial V_k} {\partial v_k}} \right)} 
{\frac {\partial v_k} {\partial \varphi}} + 
\lambda_k \delta_p(\theta) 
{\frac {\partial V_k} {\partial \varphi}} = 0, 
\end{equation} 
\noindent 
where 
\begin{eqnarray}\label{eq:potential} 
V_k{\left( \varphi, v_k; \theta \right)} = 2 {\sqrt{2}} 
\pi \int d \varphi' d \xi' \varrho_{3-k}{\left( 
\varphi', \xi'; \theta \right)} \nonumber \\ 
\times {\left| \sqrt{v_k(\varphi, \xi; \theta)} \cos \varphi - 
\sqrt{v_{3-k}(\varphi', \xi'; \theta)} \cos \varphi' \right|}. 
\end{eqnarray} 
\noindent 
It is important to note that Eqs. (\ref{eq:hydrho}) and 
(\ref{eq:hydvel}) comprise a closed set, that is (as can be easily 
checked) equations for higher moments can be reduced to these two 
equations. 

At this point we make the important conjecture that Eqs. 
(\ref{eq:hydrho}) and (\ref{eq:hydvel}) possess a stationary solution 
that is independent of the angle variable $\varphi$. Without loss of 
generality we choose 
\begin{equation}\label{eq:stat} 
v_k^{(0)} = \xi = {\rm const}, \qquad \qquad 
\varrho_k^{(0)} = G(\xi) = {\rm const}. 
\end{equation} 

\renewcommand{\theequation}{\thesection.\arabic{equation}}

\setcounter{equation}{0}

\section{SOLUTION OF THE LINEARIZED EQUATIONS}

Expressing $\varrho_k = \varrho_k^{(0)} + \varrho_k^{(1)}$ and 
$v_k = v_k^{(0)} + v_k^{(1)}$, the linearized hydrodynamic equations 
can be written as 
\begin{equation}\label{eq:linrho} 
{\frac {\partial \varrho_k^{(1)}} {\partial \theta}} + 
{\widetilde{\nu}}_k {\frac {\partial \varrho_k^{(1)}} 
{\partial \varphi}} + \lambda_k \delta_p(\theta) 
\varrho_k^{(0)} {\frac {\partial^2 V_k^{(1)}} 
{\partial \varphi \partial v_k}} = 0, 
\end{equation} 
\begin{equation}\label{eq:linvel} 
{\frac {\partial v_k^{(1)}} {\partial \theta}} + 
{\widetilde{\nu}}_k 
{\frac {\partial v_k^{(1)}} {\partial \varphi}} + 
\lambda_k \delta_p(\theta) 
{\frac {\partial V_k^{(1)}} {\partial \varphi}} = 0. 
\end{equation} 
\noindent 
Here ${\widetilde{\nu}}_k$ is the incoherently perturbed betatron 
tune, defined by 
\begin{equation}\label{eq:tune} 
{\widetilde{\nu}}_k = \nu_k + {\frac {\lambda_k} 
{2 \pi}} {\left \langle 
{\frac {\partial V_k^{(0)}} {\partial v_k}} 
\right \rangle}_{\varphi}, 
\end{equation} 
\noindent 
where the angular bracket implies an average over the angle variable. 
Next we determine the derivatives of the first-order beam-beam 
potential $V_k^{(1)}$ entering the linearized hydrodynamic equations 
corresponding to 
\begin{eqnarray}\label{eq:deriv1} 
{\frac {\partial V_k^{(1)}} {\partial \varphi}} = 
- 2 \pi {\sqrt{2 \xi}} \sin \varphi \int 
d \varphi' d \xi' \varrho_{3-k}^{(1)} {\left( 
\varphi', \xi'; \theta \right)} \nonumber \\ 
\times {\rm sgn} {\left( \sqrt{\xi} \cos \varphi - 
\sqrt{\xi'} \cos \varphi' \right)}, 
\end{eqnarray} 
\begin{eqnarray}\label{eq:deriv2} 
{\frac {\partial^2 V_k^{(1)}} {\partial \varphi 
\partial v_k}} = - \pi {\sqrt{\frac {2} {\xi}}} 
\sin \varphi \int d \varphi' d \xi' \varrho_{3-k}^{(1)} 
{\left( \varphi', \xi'; \theta \right)} \nonumber \\ 
\times {\rm sgn} {\left( \sqrt{\xi} \cos \varphi - 
\sqrt{\xi'} \cos \varphi' \right)}. 
\end{eqnarray} 
\noindent 
Finally, we obtain the linearized equation for the beam density 
\begin{eqnarray}\label{eq:linearize} 
{\frac {\partial \varrho_k^{(1)}} {\partial \theta}} + 
{\widetilde{\nu}}_k {\frac {\partial \varrho_k^{(1)}} 
{\partial \varphi}} - \pi \lambda_k \delta_p(\theta) 
\nonumber \\ 
\times {\sqrt{\frac {2} {\xi}}} \varrho_k^{(0)}(\xi) 
\sin \varphi \int d \varphi' d \xi' \varrho_{3-k}^{(1)} 
{\left( \varphi', \xi'; \theta \right)} \nonumber \\ 
\times {\rm sgn} {\left( \sqrt{\xi} \cos \varphi - 
\sqrt{\xi'} \cos \varphi' \right)} = 0. 
\end{eqnarray} 
\noindent 
In order to solve Eq. (\ref{eq:linearize}), we note that the function 
$\varrho_k^{(1)}$ may be represented as 
\begin{equation}\label{eq:repres} 
\varrho_k^{(1)}(\varphi, \xi; \theta) = 
{\frac {\varrho_k^{(0)}(\xi)} {\sqrt{\xi}}} 
{\cal R}_k(\varphi, \xi; \theta). 
\end{equation} 
\noindent 
Assuming the function $G(\xi)$ in Eq. (\ref{eq:stat}) to be of the 
form 
\begin{equation}\label{eq:gauss} 
G(\xi) = {\frac {1} {2 \pi \sigma_k^2}} 
{\exp \! {\left( - {\frac {\xi} {\sigma_k^2}} \right)}} 
\end{equation} 
\noindent 
for small vertical beam sizes $\sigma_k$, we obtain

\begin{eqnarray} 
{\frac {\varrho_k^{(0)}(\xi) 
\varrho_{3-k}^{(0)}(\xi')} 
{\sqrt{\xi \xi'}}} = {\frac {\exp \! {\left( - 
{\frac {\xi} {\sigma_k^2}} - {\frac {\xi'} 
{\sigma_{3-k}^2}} \right)}} 
{(2 \pi)^2 \sigma_k^2 \sigma_{3-k}^2 
{\sqrt{\xi \xi'}}}} \nonumber 
\end{eqnarray} 
\begin{eqnarray} 
= {\frac {\exp \! {\left( - {\frac {\xi'} 
{\sigma_{3-k}^2}} + {\frac {\xi'} 
{\sigma_k^2}} - 
{\frac {2 {\sqrt{\xi \xi'}}} 
{\sigma_k^2}} \right)}} {(2 \pi)^2 \sigma_k^2 
\sigma_{3-k}^2 
{\sqrt{\xi \xi'}}}} {\exp \! {\left[ - {\frac 
{{\left( {\sqrt{\xi}} - {\sqrt{\xi'}} \right)}^2} 
{\sigma_k^2}} \right]}} \nonumber 
\end{eqnarray} 
\begin{eqnarray}\label{eq:approx} 
\sim {\sqrt{\pi}} \sigma_k {\frac 
{\varrho_k^{(0)}(\xi) \varrho_{3-k}^{(0)}(\xi')} 
{\sqrt{\xi \xi'}}} \delta {\left( \sqrt{\xi} - 
\sqrt{\xi'} \right)}. 
\end{eqnarray} 
\noindent 
If ${\cal R}_k$ does not depend on the Lagrange variable $\xi$, making 
use of Eq. (\ref{eq:approx}), we rewrite Eq. (\ref{eq:linearize}) as 
\begin{eqnarray}\label{eq:amplit} 
{\frac {\partial {\cal R}_k} {\partial \theta}} + 
{\widetilde{\nu}}_k {\frac {\partial {\cal R}_k} 
{\partial \varphi}} - \pi {\widetilde{\lambda}}_k 
\delta_p(\theta) \sin \varphi \nonumber \\ 
\times \int d \varphi' {\cal R}_{3-k} 
{\left( \varphi'; \theta \right)} 
{\rm sgn} {\left( \cos \varphi - 
\cos \varphi' \right)} = 0, 
\end{eqnarray} 
\noindent 
where 
\begin{equation}\label{eq:lamtilde} 
{\widetilde{\lambda}}_k = {\sqrt{\frac {2} {\pi}}} 
\lambda_k {\frac {\sigma_k} {\sigma_{3-k} \Sigma}}, 
\qquad \qquad 
\Sigma = {\sqrt{\sigma_k^2 + \sigma_{3-k}^2}}. 
\end{equation} 
\noindent 
Note that this approximation is valid if and only if the perturbed 
betatron tunes in Eq. (\ref{eq:tune}) do not depend on $\xi$, which 
in general is not the case. This leads to an effect similar to 
Landau damping, well-known in plasma physics, which we shall 
neglect in what follows. Fourier transforming Eq. (\ref{eq:amplit}) 
yields 
\begin{eqnarray}\label{eq:ampfour} 
{\frac {\partial {\widetilde{\cal R}}_k(n)} 
{\partial \theta}} + i n {\widetilde{\nu}}_k 
{\widetilde{\cal R}}_k(n) \nonumber \\ 
- {\frac {{\widetilde{\lambda}}_k} {2}} 
\delta_p(\theta) \sum \limits_{m=- \infty}^{\infty} 
{\cal M}_{nm} {\widetilde{\cal R}}_{3-k}(m) = 0, 
\end{eqnarray} 
\noindent 
where 
\begin{equation}\label{eq:fourier} 
{\widetilde{\cal R}}_k(n; \theta) = 
{\frac {1} {2 \pi}} \int \limits_0^{2 \pi} 
d \varphi {\cal R}_k(\varphi; \theta) 
\exp \! (- i n \varphi), 
\end{equation} 
\begin{eqnarray}\label{eq:matrix} 
{\cal M}_{nm} = \int \limits_0^{2 \pi} d \varphi 
\int \limits_0^{2 \pi} d \varphi' e^{- i n \varphi} 
\sin \varphi e^{i m \varphi'} \nonumber \\ 
\times {\rm sgn} {\left( \cos \varphi - 
\cos \varphi' \right)}. 
\end{eqnarray} 
\noindent 
In order to determine the infinite matrix ${\cal M}$, we utilize the 
integral representation of the sign-function 
\begin{equation}\label{eq:sign} 
{\rm sgn}(x) = {\frac {1} {\pi i}} \int 
\limits_{- \infty}^{\infty} 
{\frac {d \lambda} {\lambda}} \exp \! (i \lambda x). 
\end{equation} 
\noindent 
As a result, we obtain 
\begin{eqnarray}\label{eq:elem} 
{\cal M}_{nm} = 4 \pi n i^{n-m+1} \int 
\limits_{- \infty}^{\infty} 
{\frac {d \lambda} {\lambda^2}} 
{\cal J}_m(\lambda) {\cal J}_n(\lambda) 
\nonumber \\ 
= \left\{ \begin{array}{ll}
- {\frac {32 i n} {{\left[ (n+m)^2 - 1 \right]} 
{\left[ (n-m)^2 - 1 \right]}}}, & 
\mbox{for $n+m=$ even}, \\ 
0, & \mbox{for $n+m=$ odd}, 
\end{array} 
\right. 
\end{eqnarray} 
\begin{eqnarray}\label{eq:eleme} 
{\cal M}_{mn} = (-1)^{m-n} {\frac {m} {n}} 
{\cal M}_{nm}, 
\end{eqnarray} 
\noindent 
where use has been made of 
\begin{eqnarray}\label{eq:bessel} 
e^{i z \cos \varphi} = \sum 
\limits_{n=- \infty}^{\infty} i^n {\cal J}_n(z) 
e^{i n \varphi}, 
\nonumber \\ 
{\cal J}_{n-1}(z) + {\cal J}_{n+1}(z) = 
{\frac {2 n} {z}} {\cal J}_n(z). 
\end{eqnarray} 
\noindent 
Here ${\cal J}_n(z)$ is the Bessel function of the first kind of 
order $n$. 

\renewcommand{\theequation}{\thesection.\arabic{equation}}

\setcounter{equation}{0}

\section{COHERENT BEAM-BEAM RESONANCES}

Equation (\ref{eq:ampfour}) can be formally solved to obtain the 
one-turn transfer map 
\begin{eqnarray}\label{eq:map} 
{\widetilde{\cal R}}_k (n; 2 \pi) = 
\exp \! {\left( -  2 \pi i n {\widetilde{\nu}}_k \right)} 
\nonumber \\
\times {\left[ {\widetilde{\cal R}}_k (n; 0) + 
{\frac {{\widetilde{\lambda}}_k} {2}} \sum 
\limits_{m=- \infty}^{\infty} {\cal M}_{nm} 
{\widetilde{\cal R}}_{3-k} (m; 0) \right]}. 
\end{eqnarray} 
\noindent 
Consider now a coherent beam-beam resonance of the form 
\begin{equation}\label{eq:reson} 
n_1 {\widetilde{\nu}}_1 + 
n_2 {\widetilde{\nu}}_2 = s + \Delta, 
\end{equation} 
\noindent 
where $n_1$, $n_2$ and $s$ are integers, and $\Delta$ is the resonance 
detuning. Retaining only the $\pm n_1$ and the $\pm n_2$ elements in 
${\cal M}_{nm}$, the transformation matrix of the coupled map 
equations (\ref{eq:map}) can be expressed as 
\begin{eqnarray}\label{eq:transfer} 
{\left( \begin{array}{clcr} 
e^{- i \psi_1} & 0 & \alpha_1 e^{- i \psi_1} & 
\alpha_1 e^{- i \psi_1} \\ 
0 & e^{i \psi_1} & - \alpha_1 e^{i \psi_1} & 
- \alpha_1 e^{i \psi_1} \\ 
\alpha_2 e^{- i \psi_2} & \alpha_2 e^{- i \psi_2} & 
e^{- i \psi_2} & 0 \\ 
- \alpha_2 e^{i \psi_2} & - \alpha_2 e^{i \psi_2} & 
0 & e^{i \psi_2} 
\end{array} 
\right)}, 
\end{eqnarray} 
\noindent 
where 
\begin{equation}\label{eq:psi} 
\psi_k = 2 \pi n_k {\widetilde{\nu}}_k, 
\qquad \qquad 
\alpha_1 = {\frac {{\widetilde{\lambda}}_1} {2}} 
{\cal M}_{n_1 n_2}, 
\end{equation} 
\begin{equation}\label{eq:alpha} 
\alpha_2 = {\frac {{\widetilde{\lambda}}_2} {2}} 
(-1)^{n_2-n_1} {\frac {n_2} {n_1}} 
{\cal M}_{n_1 n_2}. 
\end{equation} 
\noindent 
The eigenvalues of the transfer matrix defined in Eq. 
(\ref{eq:transfer}) are the roots of the secular equation
\begin{eqnarray}\label{eq:secular} 
{\left( \lambda^2 - 2 \lambda \cos \psi_1 + 1 
\right)}
{\left( \lambda^2 - 2 \lambda \cos \psi_2 + 1 
\right)} + \nonumber \\ 
+ 2 \alpha_1 \alpha_2 {\left[ 
\cos {\left( \psi_1 - \psi_2 \right)} - 
\cos {\left( \psi_1 + \psi_2 \right)} 
\right]} \lambda^2 = 0. 
\end{eqnarray} 
\noindent 
Casting Eq. (\ref{eq:secular}) in the form 
\begin{eqnarray}\label{eq:secul} 
{\left( \lambda^2 - 2 c_1 \lambda + 1 \right)}
{\left( \lambda^2 - 2 c_2 \lambda + 1 \right)} = 0, 
\end{eqnarray} 
\noindent 
where 
\begin{eqnarray}\label{eq:coeff} 
c_{1,2} = {\frac {\cos \psi_1 + \cos \psi_2} {2}} 
\pm \nonumber \\ 
\pm {\frac {1} {2}} {\sqrt{{\left( 
\cos \psi_1 - \cos \psi_2 \right)}^2 - 
4 A \sin \psi_1 \sin \psi_2}}, 
\end{eqnarray} 
\begin{equation}\label{eq:coeffa} 
A = {\frac {{\widetilde{\lambda}}_1 
{\widetilde{\lambda}}_2} {4}} 
(-1)^{n_2-n_1} {\frac {n_2} {n_1}} 
{\cal M}_{n_1 n_2}^2, 
\end{equation} 
\noindent 
we obtain the stability criterion 
\begin{equation}\label{eq:stability} 
{\left| \cos \psi_1 \cos \psi_2 + 
A \sin \psi_1 \sin \psi_2 \right|} 
< 1. 
\end{equation} 

To conclude this section we note that in the case of a symmetric 
collider the stopbands calculated from Eq. (\ref{eq:stability}) 
coincide with the results obtained by Chao and Ruth [see Eq. (31) 
of Ref. 1]. 

\renewcommand{\theequation}{\thesection.\arabic{equation}}

\setcounter{equation}{0}

\section{CONCLUDING REMARKS}

Based on the Radon transform we have developed a macroscopic fluid 
model of the coherent beam-beam interaction. The linearized 
hydrodynamic equations are further solved and a stability criterion 
for coherent beam-beam resonances have been found in closed form. 

\section{ACKNOWLEDGMENTS}

It is a pleasure to thank Prof. A. Chao and Dr. Y. Cai for helpful 
discussions concerning the subject of the present paper. This 
research was supported by the U.S. Department of Energy.


\begin{thebibliography}{9} 

\bibitem{chao} A.W. Chao and R.D. Ruth, {\it Particle Accelerators}, 
{\bf 16} 201 (1985). 

\bibitem{deans} S.R. Deans, ``{\it The Radon Transform and Some of 
Its Applications}'', Wiley, New York 1992.

\bibitem{tzenov} Stephan I. Tzenov, {\it FERMILAB-Pub-98/275}, 
Batavia 1998. 

\end{thebibliography}
\end{document}